\documentclass[doublecol]{epl2}
\usepackage{amsfonts,amssymb,amsmath,color}
\usepackage{graphicx}
\usepackage{bm}

\begin{document}

\title{Power-laws in recurrence networks from dynamical systems}
\author{Y.~Zou \inst{1,2,3} \and J.~Heitzig \inst{2} \and R.\,V.~Donner \inst{2} \and J.\,F.~Donges \inst{2,5} \and J.\,D.~Farmer \inst{2,4} \and R.~Meucci \inst{6} \and S.~Euzzor \inst{6} \and N.~Marwan \inst{2} \and J.~Kurths \inst{2,5,7} }
\shortauthor{Y.~Zou \etal}
\institute{
\inst{1} Department of Physics, East China Normal University, Shanghai, China \\
\inst{2} Potsdam Institute for Climate Impact Research, Potsdam, Germany\\
\inst{3} Department of Electronic and Information Engineering, The Hong Kong Polytechnic University, Kowloon, Hong Kong \\
\inst{4} Santa Fe Institute, Santa Fe, New Mexico, USA \\
\inst{5} Department of Physics, Humboldt University, Berlin, Germany \\
\inst{6} Department of Physics, University of Florence, Florence, Italy \\
\inst{7} Institute for Complex Systems and Mathematical Biology, University of Aberdeen, Aberdeen, United Kingdom
}



\abstract{
Recurrence networks are a novel tool of nonlinear time series analysis allowing
the characterisation of higher-order geometric properties of complex dynamical
systems based on recurrences in phase space, which are a fundamental concept in
classical mechanics. In this Letter, we demonstrate that
recurrence networks obtained from various deterministic model systems as well as
experimental data naturally display power-law degree distributions with scaling
exponents $\gamma$ that can be derived exclusively from the systems' invariant
densities. For one-dimensional maps, we show analytically that $\gamma$ is not
related to the fractal dimension. For continuous systems, we find two distinct
types of behaviour: power-laws with an exponent $\gamma$ depending on a suitable
notion of local dimension, and such with fixed $\gamma=1$.
}

\maketitle

\def\averagek{\langle k\rangle}

\section{Introduction}

Power-law distributions have been widely observed in diverse fields
such as seismology, economy, and finance in the context of critical
phenomena~\cite{Gutenberg1944,Mantegna1995,Farmer2004}. In many cases, the underlying complex systems can
be regarded as networks of mutually interacting subsystems with a complex
structural organisation. Specifically, numerous examples have been found for
hierarchical structures in the connectivity of such complex networks, \emph{i.e.,} the presence of scale-free distributions $P(k)\sim k^{-\gamma}$ of the node degrees~\cite{Albert2002,Newman2003}. Such hierarchical organisation is particularly well expressed in network of networks, or interdependent networks, which constitute an emerging and important new field of complex network research~\cite{Gao2012,donges2011d}. The interrelationships between the non-trivial structural properties of complex networks and the resulting dynamics of the mutually interacting
subsystems are subject of intensive research~\cite{Boccaletti2006,Arenas2008}.

\begin{figure}
  \centering
  \includegraphics[width=0.9\columnwidth]{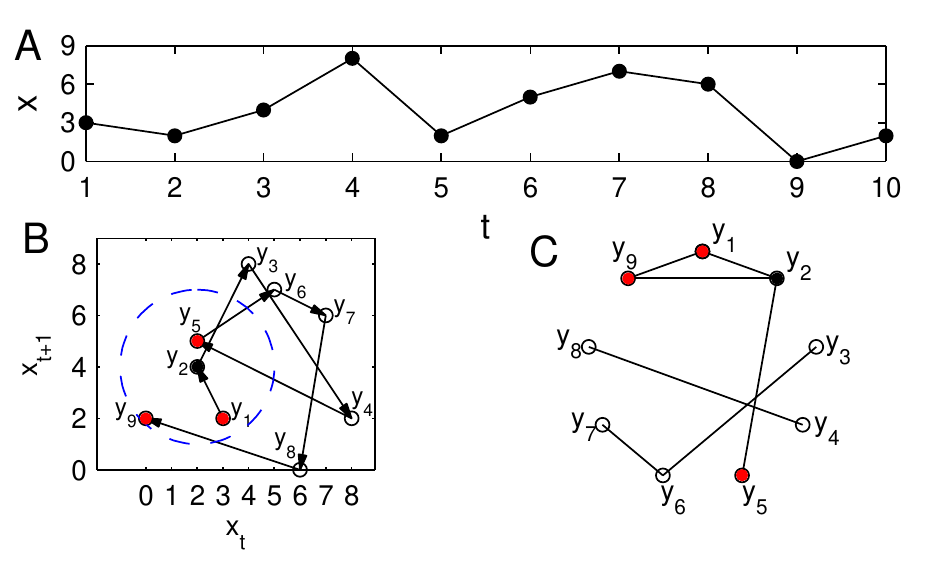}
  \caption{\small {(Colour online) Construction of a RN:  (A)
      time series $x_t$, (B) reconstructed two-dimensional phase space
      trajectory with state vectors $\mathbf{y}_t = (x_t,x_{t+1})$ and an
      exemplary $\varepsilon$-ball around $y_2$ (dashed) for
      $\varepsilon=3$.  (C) Resulting recurrence network.
  \label{ts_net}}}
\end{figure}

Among other developments, one of the main recent achievements of complex network
theory are various conceptionally different approaches for statistically
characterising dynamical systems by graph-theoretical
methods~\cite{Zhang2006,Lacasa2008,Xu2008,Marwan2009}. In this Letter, we report and thoroughly
explain the emergence of power-laws in the degree distribution of so-called
\textit{recurrence networks} (RNs)~\cite{Marwan2009,Donner2010NJP,Donner2010PRE,Donges2011PNAS} for various paradigmatic
model systems as well as experimental data. RNs encode the underlying system's
\textit{recurrences in phase space} and are based on a fundamental concept in
classical physics~\cite{Poincare1890}. Due to their direct link to dynamical systems theory, RNs are
probably the most widely applicable type of complex networks inferred from time
series introduced so far. Although the system's temporal evolution
cannot be reconstructed from the RN, this representation allows for an analysis
of the attractor's geometry in phase space using techniques from network theory. Specifically, nodes represent individual state
vectors, and pairs of nodes are linked when they are mutually closer than some
threshold distance $\varepsilon>0$~\cite{Marwan2007} (cf.\,Fig.\,\ref{ts_net}).
According to this definition, RNs are random geometric
graphs~\cite{Herrmann2003} (\emph{i.e.}, undirected {\em spatial
networks}~\cite{Barthelemy2011}), where the spatial distribution of nodes is
completely determined by the probability density function of the invariant
measure of the dynamical system under study, and links are established according
to the distance in phase space. Consequently, their degree distribution $P(k)$
directly relates to the system's invariant density $p(x)$.

In this work, we demonstrate the emergence of scaling in the degree
distributions of RNs and provide some evidence that this phenomenon is (unlike
many other scaling exponents occurring in the context of dynamical systems)
commonly unrelated to the fractal attractor dimension, except for some
interesting special cases. Instead, the power-laws naturally arise from the
variability of the invariant density $p(x)$ of the system (\emph{i.e.,} peaks or
singularities of $p$), as we will show numerically as well as explain theoretically.

\section{Power-law scaling and singularities of the invariant density}

\begin{figure}[htb]
  \centering \includegraphics[width=\columnwidth]{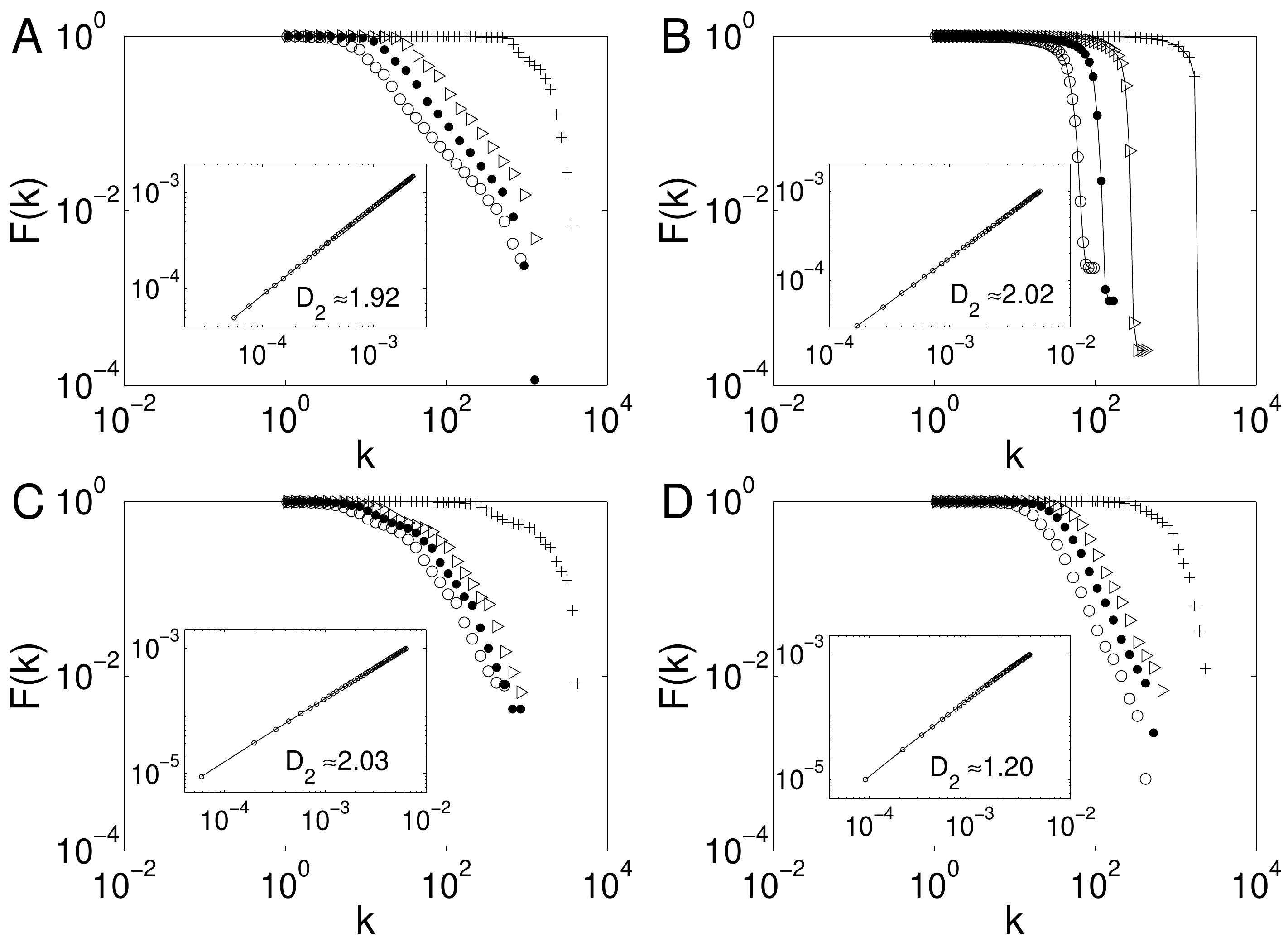}
  \caption{\small {Cumulative degree distributions
  $F(k)=\sum_{k'=k}^{\infty} P(k')$ of the RNs obtained for several discrete maps: (A) $x$-component of the first return map of the R\"ossler system ($\gamma=2.16 \pm 0.03$), (B) map of consecutive local maxima of the $z$-component of the Lorenz system with $r=28$ (no scaling), (C) as in (B) with $r=90$ ($\gamma = 2.64 \pm 0.18$), and (D) H\'enon map ($\gamma = 2.88 \pm 0.04$). Estimates of $\gamma>1$ have been obtained from the cumulative degree distributions $F(k) \sim k^{1-\gamma}$ (in the following figures, we show either $F(k)$ or $P(k)$ when sufficiently straight in a log-log plot) by means of a maximum likelihood approach~\cite{Clauset2009} as averages considering 100 different values of $\rho$ for which a power-law appears. For each $\rho$, $5$ different realisations with random initial conditions are used. We consider RNs of size $N=2\times 10^5$ and the Euclidean norm in all examples discussed in this Letter. Four cases were chosen for illustration corresponding to a link density of $\rho_1 = 0.02 \%\ (\circ)$, $\rho_2 = 0.03 \%\ (\bullet)$, $\rho_3 = 0.05 \%\ (\triangleright)$, and $\rho_4 = 3 \%\ (+)$. A power-law is hardly detectable for $\rho_{4}$. Insets: corresponding log-log plot of the correlation sum $C(\varepsilon)$ vs.\ $\varepsilon$, where the correlation dimension $D_2$ is estimated by linear regression and agrees well with values from the  literature~\cite{Grassberger1983PRL,Sprott2003,Kantz1997}.
      \label{deg_rlh}}}
\end{figure}

As initial examples, Fig.\,\ref{deg_rlh} illustrates the presence of power-law degree distributions in the RNs obtained for several prototypical low-dimensional chaotic systems with a suitable choice of the systems' characteristic parameters: (i) the R\"ossler system in spiral-chaos regime: $\dot{x} =-y - z$, $\dot{y} = x + 0.2y$, $\dot{z} = 0.2 + z (x - 5.7)$; (ii) the Lorenz system: $\dot{x} = 10 ( y - x )$, $\dot{y} = x ( r - z ) - y$, $\dot{z} = x y - 8/3 z$; and (iii) the H\'enon map: $x_{n+1}  = 1 - 1.4x_{n}^{2} + y_{n}$, $y_{n+1}  = 0.3x_{n}$. For the R\"ossler and Lorenz systems, a proper time discretisation has been used as explained in detail in the figure caption. 

In all cases, scaling emerges only if the distance threshold $\varepsilon$ is chosen
small enough, which corresponds to a small average degree $\averagek$ and link
density $\rho=\averagek/(N-1)$ of the resulting network ($N$ being the number of
nodes). We note that the respective range of $\varepsilon$ should be
sufficiently higher than the threshold for which a giant component exists
\cite{Donges2012}. The size of the scaling regime decreases with growing $\rho$
and becomes hardly detectable for  $\rho\gtrsim 1\%$ (Fig.\,\ref{deg_rlh}). The
distance threshold $\varepsilon$ also occurs in dimension estimation where the
limit $\varepsilon\to 0$ is taken (\emph{e.g.,} \cite{Farmer1983}). In contrast,
a RN is based on one finite $\varepsilon$.  Our smallest $\varepsilon$ are,
however, still large enough to avoid the problem of lack of neighbours~\cite{Sprott2003} since the correlation integral $C(\varepsilon)$ still shows the same dependency on $\varepsilon$ as for larger values (see insets in Fig.\,\ref{deg_rlh}).
  
A theoretical explanation of the emergence of power-law distributions of
RN-based degree is based on the general theory of random geometric
graphs~\cite{Herrmann2003}, where nodes are sampled from some probability density function $p(x)$. For a RN, the space is the phase space of a dynamical system and the nodes are states sampled at discrete times.  If we assume that the system is ergodic, the sampled trajectory is already close to its attractor, and the sampling times are generic (particularly, the sampling interval is co-prime to any period lengths of the system), the nodes can be interpreted as being sampled from the probability density function $p(x)$ of the invariant measure $\mu$ of the attractor~\cite{Eckmann1985}. The degree distribution of a general random geometric graph, $P(k)$, is derived from $p(x)$ in the limit of large network size $N$ as~\cite{Herrmann2003} 
\begin{equation}\label{herrmann}
  P(k) = \int dx\,p(x) e^{-\alpha p(x)}(\alpha p(x))^k/k!
\end{equation}
with $\alpha = \averagek / \int dx\,p(x)^2$ (note that the computation of $\averagek$ involves integration of $p(x)$ over the $\varepsilon$-neighbourhood of all points $x$ and thus implicitly depends on the specifically chosen $\varepsilon$ as well as the sample size $N$). Hence, the invariant density of the system exclusively determines the existence of a power-law in $P(k)$  and its exponent $\gamma$.

For systems with a {\em one-dimensional phase space,} it can be shown that under
some weak conditions on $p(x)$ it holds:
\begin{equation} 
\textstyle P(k) \approx
  \frac{k+1}{\alpha}\sum_{x\in
    p^{-1}(\frac{k+1}{\alpha})}|p'(x)|^{-1}.\label{eq:pk}
\end{equation}
This implies that if $p(x)$ has a power-law-shaped peak at some state $x_0$,
\emph{i.e.,} $p(x)\sim|x-x_0|^{-1/\gamma}$ for some $\gamma>0$, the degree
distribution $P(k)$ also follows a power-law but with the reciprocal exponent,
$P(k)\sim k^{-\gamma}$. Specifically, a slower decaying invariant density leads
to a faster decaying degree distribution. Note that not all invariant densities
lead to power-laws: If $p(x)$ is Gaussian, we get $P(k) \approx 2(-2 \ln {k
\sqrt{2 \pi}/\alpha})^{-1/2}$ instead of a power-law.

More generally, we can deduce that the presence of singularities in the invariant density is the key feature determining whether or not the resulting RN has a power-law degree distribution. This relationship can be intuitively understood: If $p(x)$ has a singularity at some point $x_0$ in phase space, then a time series of the associated dynamical system will return very often to the neighbourhood of $x_0$. Hence, nodes with a high degree will accumulate close to the singularity. If the resulting invariant density obeys a power-law decay, Eq.~(\ref{eq:pk}) implies the emergence of a power-law degree distribution. If there is more than one singularity of $p(x)$, one can expect the resulting degree distribution being related to a weighted sum of the influences of these points. Vice versa, the presence of a power-law degree distribution in the RN of a dynamical system requires the existence of a power-law in the invariant density, \emph{i.e.,} the presence of a singularity. We therefore conjecture that a local power-law in the invariant density is a necessary and sufficient condition for the emergence of a scale-free RN. Beyond the explicit results for one-dimensional systems as discussed above, we further show below that the emergence of a power-law scaling is also possible in higher dimensions and conjecture that this requires the presence of a dynamically invariant object (\emph{e.g.,} an unstable or hyperbolic fixed point) close to which the invariant density scales as a power-law at least in one direction.



\section{Power-law scaling vs. fractal dimension}

\begin{figure}[t]
  \centering
  \includegraphics[width=\columnwidth]{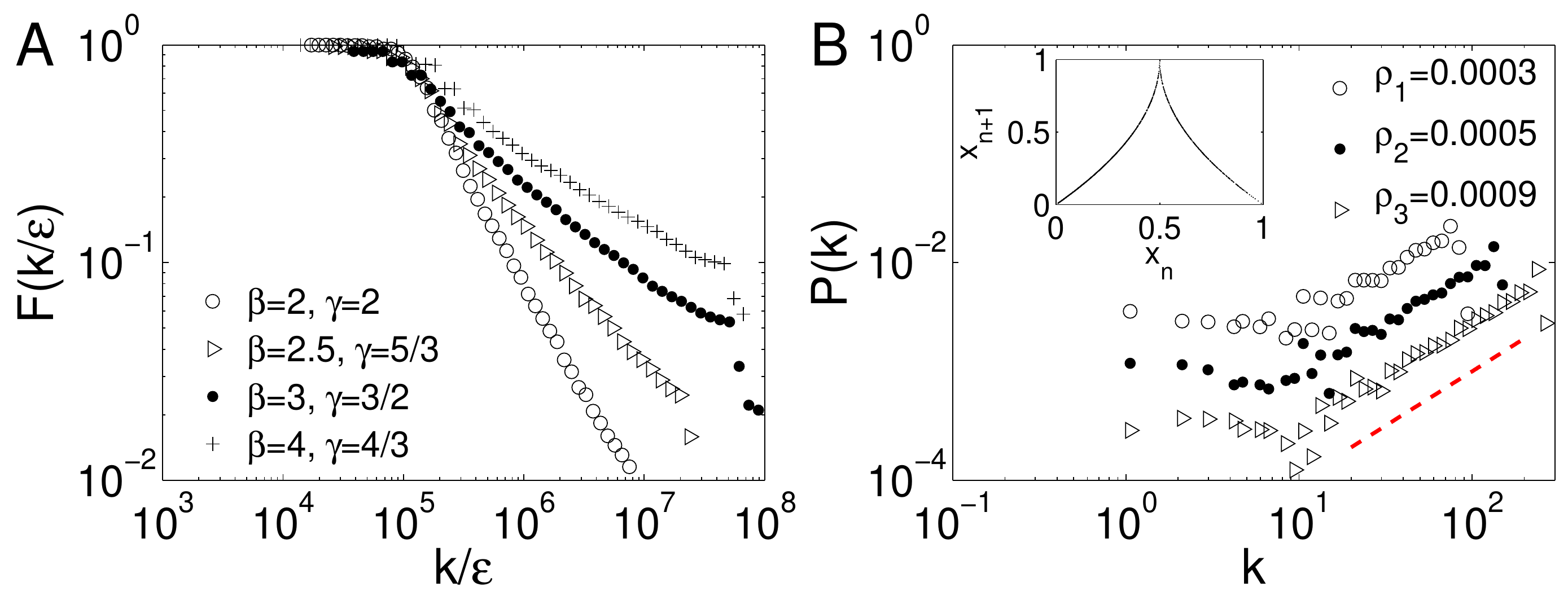}
  \caption{\small {(Colour online) (A): Cumulative degree distribution
      $F(k)$ of the RNs for the map (\ref{map_oneD}) with $\beta=2, 2.5, 3, 4$.
      Note that we divide the degree $k$ by $\varepsilon$
      so that the $x$-axis is $\varepsilon$-dimensionless.  (B) $P(k)$ of
      the cusp map $\beta=0.5$. The inset shows the corresponding
      iterative function $f(x)$. The dashed line of (B) has a slope of
      $1$. The legend indicates the respective link densities $\rho$. }
    \label{deg_betaAll} }
\end{figure}

As an example for {\em discrete-time systems} leading to power-laws, consider
the generalised logistic map~\cite{Lyra1998}
\begin{equation} \label{map_oneD} 
f(x) = 1 - |2x - 1|^{\beta}
\end{equation}
with $\beta \geqslant \frac{1}{2}$. For $\beta = \frac{1}{2},1,2$, this gives
the cusp map, tent map, and standard logistic map, respectively. For general
$\beta>0$, the unit interval $[0,1]$ is mapped onto itself by a symmetric
function with a maximum of $1$ at $x=\frac{1}{2}$, thus having two pre-images
for each $x<1$. For $\beta>1$, the associated invariant density $p(x)$ has two
peaks at $x=0$ and $x=1$ with $p(\delta)=p(1-\delta) \sim
\delta^{(1-\beta)/\beta}$ for small $\delta$. Hence, the degree distribution
$P(k)$ shows a power-law with the exponent
\begin{equation}\label{gamma_beta}
\textstyle \gamma = \beta / (\beta - 1). 
\end{equation}
Numerical results shown in Fig.\,\ref{deg_betaAll}A for several different
values of $\beta$ agree precisely with Eq.~(\ref{gamma_beta}).
In contrast, for $\beta=1$ the nodes are uniformly distributed, and the degree
distribution derived from Eq.\,(\ref{herrmann}) is Poissonian, $P(k) =
e^{-\alpha}\alpha^k/k!$ \cite{Dall2002}. For $\beta=\frac 1 2$, we get $p(x) = 2 - 2x$ \cite{Gyoergyi1984,Hemmer1984}, which leads to a specific type of ``power-law'' in $P(k)$ with $\gamma = -1$ as shown
in Fig.\,\ref{deg_betaAll}B. These results imply that the scaling exponent is
\textit{not} simply related to the fractal dimension: the attractor has the 
box-counting dimension $D_0=1$ independently of $\beta$, whereas $\gamma$ changes with
varying $\beta$ (Eq.~(\ref{gamma_beta})). However, the correlation dimension
$D_2$ also depends on $\beta$ ($D_2=1$ for $\beta \le 2$, and $D_2 = 2/\beta$ if
$\beta > 2$~\cite{Sprott2003}), \emph{i.e.,} there is an indirect relationship
between $\gamma$ and $D_2$ for certain special cases. The different behaviour of the mentioned dimensions results from the fact that $D_0$ exclusively considers the number of boxes required for covering the attractor, but not their individual probability masses as $D_2$ and other notions of fractal dimensions do.

\begin{figure}[t]
  \centering \includegraphics[width=\columnwidth]{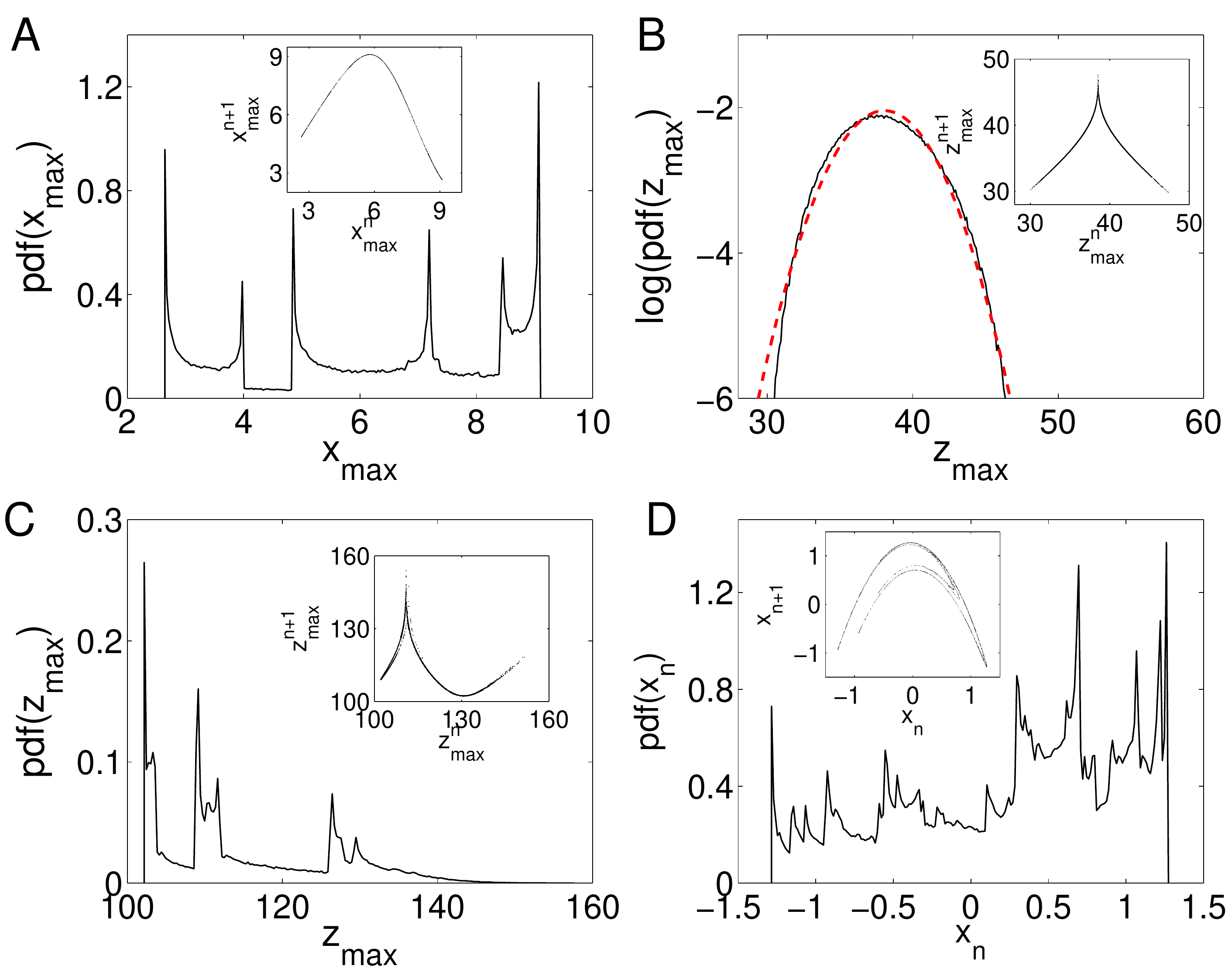}
  \caption{\small {(Colour online) PDF of (A) the successive $x$ values
      passing the plane of $y=0$ of the R\"ossler attractor, (B,C) PDF of
      $z_{max}$ for the Lorenz system of $r=28$ (B) and $r=90$ (C), and (D)
      marginal density of the $x$ component of the H\'enon map. The insets show
      the associated first return plot. The dashed curve in (B) shows a fit by a
      Gaussian distribution. }
    \label{measure_hist}}
\end{figure}

Turning to {\em continuous-time systems,} we next compare the above findings
with those for some {\em discretised} standard examples. On the
one hand, for the R\"ossler system, we consider the successive $x$-values when
passing the Poincar\'e section at $y=0$ with $\dot{y}<0$. As shown in the inset of
Fig.\,\ref{measure_hist}A, the resulting first return map has a shape similar
to the case of $\beta = 1.87$ in Eq.\,(\ref{map_oneD}). Hence, we expect a
power-law with the exponent $\gamma\approx 1.87/(1.87-1)=2.15$. The invariant
density has several dominant peaks, which are together responsible for the
power-law observed in Fig.\,\ref{deg_rlh}A with $\gamma$ indeed close to
$2.15$. In fact, $P(k)$ is a mixture of individual power-laws corresponding to the individual peaks of $p(x)$, whose exponents are all roughly the same. On the other hand, for the Lorenz system, we obtain a one-dimensional map by studying the local maxima $z_{max}^{n}$ of $z$ for
successive cycles~\cite{Lorenz_jas_1963}, \emph{i.e.,} mapping $z_{max}^{n}$ to $z_{max}^{n+1}$ (inset of Fig.\,\ref{measure_hist}B). For $r=28$, this first return map has a similar shape as Eq.\,(\ref{map_oneD}) for $\beta=0.5$ (inset of Fig.~\ref{deg_betaAll}B), but the
corresponding density is bell-shaped without a peak. Indeed, we do not observe a
power-law for $P(k)$ in this case (Fig.\,\ref{deg_rlh}B). However, increasing
$r$ changes the shape of $p(x)$ qualitatively. For example, at $r=90$
(Fig.\,\ref{measure_hist}C) the density has peaks at several points,
explaining the observed power-law in Fig.\,\ref{deg_rlh}C. A similar behaviour
can be observed for the H\'enon map, though the marginal invariant density of
the $x$-component has a more complex structure (Fig.\,\ref{measure_hist}D).

\begin{figure}[t]
  \centering
  \includegraphics[width=\columnwidth]{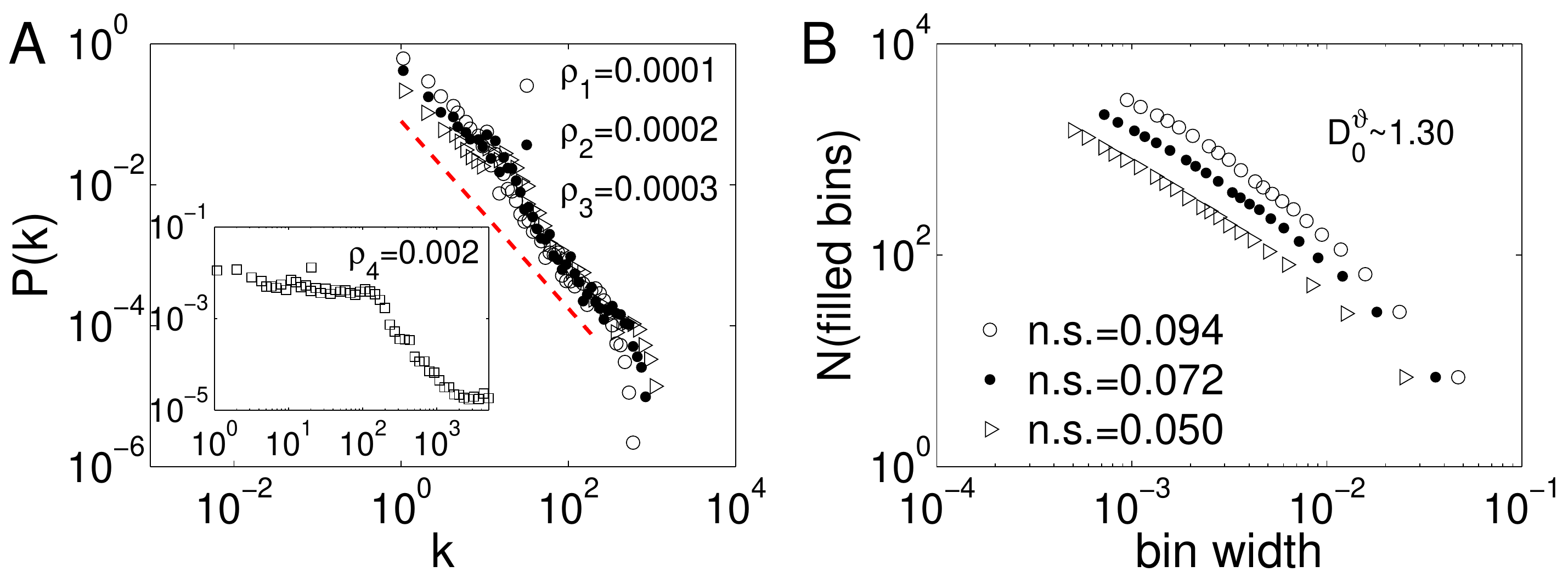}
  \caption{\small {(Colour online) (A): $P(k)$ for the RN of the R\"ossler system
      with screw-type chaos~\cite{Gaspard1983}. Three different link densities
      are chosen for illustration as indicated by the legend, a larger link
      density $\rho_4$ does not produce a power-law (inset).  The slope of the
      dashed line is $-1.33$. (B): $\vartheta$-capacity dimension
      $D_{0}^{\vartheta} \approx 1.30$ in three small cuboidal neighbourhoods of
      different size (in terms of phase space distance in each coordinate
      direction, see the legend). }
    \label{homo_ros} }
\end{figure}

\begin{figure}[t]
  \centering \includegraphics[width=\columnwidth]{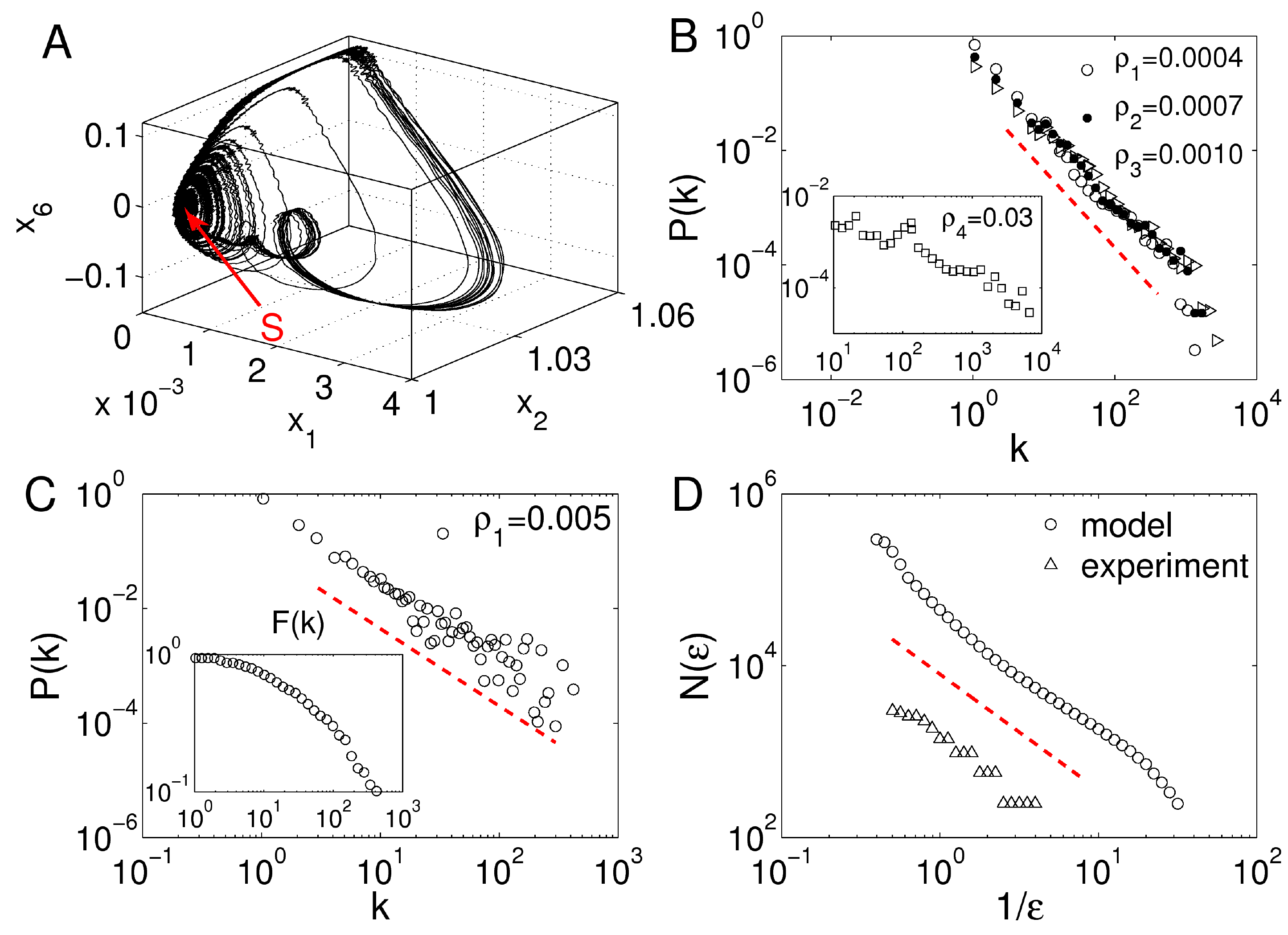}
  \caption{\small {(Colour online) Experimental laser data: (A) Phase
      portrait in subspace of $(x_1, x_2, x_6)$, where the saddle focus $S$ is
      at the most dense region. (B) $P(k)$ of RNs from model data for three
      different link densities (see the legend); a larger link density
      $\rho_{4}$ does not lead to a power-law (inset).  (C) $P(k)$ of RNs from
      experimental data (inset: cumulative distribution $F(k)$).  (D) Point-wise
      dimension $D_{1}^{S} = 1.35$. All dashed lines have slope $-1.35$.
      }
    \label{laser_degree}}
\end{figure}

\begin{figure}[t]
  \centering
  \includegraphics[width=\columnwidth]{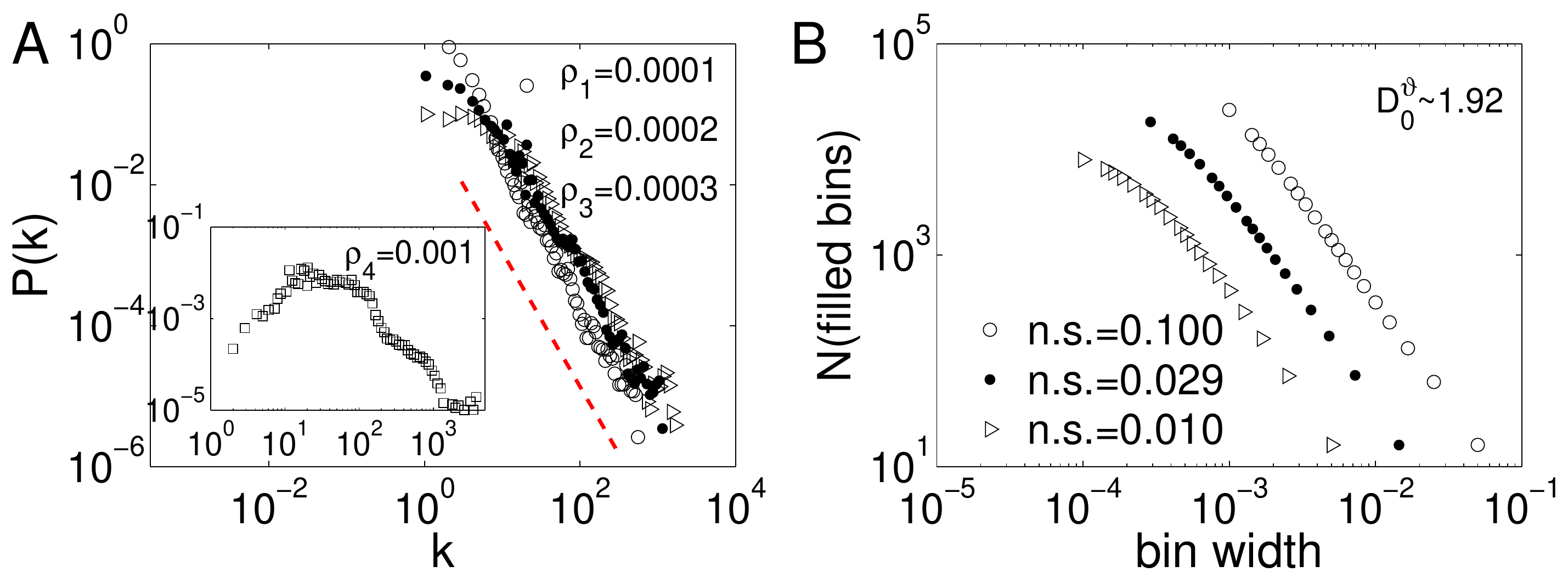}
  \caption{\small {(Colour online) As in Fig.~\ref{homo_ros} for the Lotka-Volterra system~\cite{Vano_nonlinearity2006} with homoclinic chaos. The slope of the dashed line in panel (A) is $-1.90$.}
    \label{homo_lotkavolt} }
\end{figure}

While there is no unambiguous relationship between $\gamma$ and the fractal
dimension already for discrete systems, the situation becomes even more complicated for {\emph {continuous-time}} systems which are not discretised via a Poincar\'e section or otherwise. For two-dimensional flows $\dot{x}=\Phi(x)$ with only one
peak in $p(x)$, the respective type of behaviour depends on the eigenvalues of
the Jacobian $D\Phi(x_0)$ at the fixed point $x_0$ as well as on the shape of
$p(x)$. Specifically, in many cases (that shall not be further discussed here) the existence of a power-law for $P(k)$ cannot be evaluated easily, whereas in other cases, one can analytically derive a power-law with a very small exponent of $\gamma=1$. In turn, the
following numerical results suggest that there are also examples displaying a
distinct relationship between $\gamma$ and some suitably defined local
dimension:

For the R\"ossler system in the regime of screw-type chaos with a homoclinic
point at the origin fulfilling the Shilnikov condition~\cite{Shilnikov1970,Gaspard1983}, the
invariant density is dominated by its peak at the origin. The degree
distribution $P(k)$ of the corresponding RN shows a power-law with $\gamma
\approx 1.33$, which agrees fairly well with the $\vartheta$-capacity dimension
$D_{0}^{\vartheta}$ defined in \cite{Farmer1983} (Fig.~\ref{homo_ros}). 
We also observe similar scaling laws for both numerical model and experimental
data (output intensities) of a single-mode $\text{CO}_{2}$
laser~\cite{laser_epjd2001}. The underlying system has a saddle-focus $S$
embedded in the chaotic attractor (Fig.~\ref{laser_degree}A) which causes a
spiking dynamics~\cite{Channell_prl2007,Arecchi1987}. The attractor is dominated by a
homoclinic orbit emerging from and converging to $S$. 
The degree distributions $P(k)$ resulting from both model and experimental data
suggest power-laws with $\gamma \approx 1.35$ (Fig.~\ref{laser_degree}B), which
qualitatively agrees well with the point-wise dimension of the attractor around
$S$. Finally, similar results can be obtained for a predator-prey food-chain model with four competing species~\cite{Vano_nonlinearity2006}, which also displays
homoclinic chaos, where we observe $\gamma\approx 1.9$ in agreement with
$D_{0}^{\vartheta}$ (Fig.~\ref{homo_lotkavolt}). This variety of examples underlines the general importance
and wide applicability of our findings.

\section{Technical aspects}

In general, we have to make two cautionary notes on the numerical study of
scaling laws in RNs. 

First, in the previous continuous-time examples, the
presence of power-laws with the numerically estimated exponents (see above)
cannot be rejected on a 90\% significance level using Kolmogorov-Smirnov tests.
However, the alternative of a power-law with $\gamma=1$ can also not be rejected
at the same level. Hence, power-laws with qualitatively different exponents
describe the data comparably well. It remains an open problem to determine the
correct $\gamma$. We note that this is a general problem when evaluating hypothetical power-laws from finite data \cite{Clauset2009,Stumpf2012}.

Second, experimental data often consist of only one measured
variable. Hence, a reconstruction of the associated phase space trajectory is
necessary prior to RN analysis, \emph{e.g.,} by time-delay
embedding~\cite{Packard_prl_1980}. Like estimates of dynamical invariants or
complexity measures~\cite{Letellier_pre_2005,Letellier2006}, the power-law behaviour of $P(k)$
can depend on the particular observable, because different coordinates of a
dynamical system often have different marginal densities. Specifically,
embedding theorems ensure \emph{topological} invariance (\emph{i.e.,} properties of
the dynamical system that do not change under smooth coordinate transformations
are preserved), but no \emph{metric} invariance of the attractor's geometry
including $p(x)$. For example, the logistic map and the tent map ($\beta=2, 1$ in
Eq.\,(\ref{map_oneD}), respectively) are topologically equivalent under the
transformation $x\mapsto\sin^{2}(\pi x / 2)$, but the different invariant
densities with respect to their original coordinates (that have been used for constructing the RNs from metric distances in their respective phase spaces) lead to distinct scaling exponents $\gamma$
(Fig.\,\ref{deg_betaAll}A).

\section{Conclusions}

In summary, we have reported an interesting novel aspect of the geometrical
organisation underlying the dynamics of many complex systems in physics and beyond. Specifically, we have provided an analytical explanation of the emergence of power-laws in recurrence networks constructed from sampled time series based on the theory of random geometric graphs. Unlike for comparable complex network approaches~\cite{Zhang2006,Lacasa2008}, this scaling is not simply related
to the system's fractal dimension, but determined by both the singularities of the invariant density and the considered spatial scale $\varepsilon$. We emphasise that dimensions are defined in the limit of $\varepsilon \to 0$ and practically estimated by a series of $\varepsilon$ values, whereas the power-law exponent $\gamma$ of the RN appears for each sufficiently small $\varepsilon$ \emph{individually}. Note that in contrast to the degree, the transitivity properties of RNs have a direct relationship with attractor dimension~\cite{Donner2011EPJB}.

In comparison with the invariant density itself, fractal dimensions are a rather specific characteristic. In particular, they do not simply describe the whole system (as the invariant density itself does), but quantify density variations on the attractor viewed at different spatial scales $\varepsilon$~\cite{Farmer1983,Halsey_pra_1986}. Conversely, the scaling exponent $\gamma$ directly characterises a power-law decay of the density in phase space independent of a specific scale. In this spirit, both fractal dimension and scaling exponent $\gamma$ capture conceptually different aspects of the geometric organisation of a dynamical system in its phase space. However, although there is no general relationship between $\gamma$ and fractality, in some special cases the power-law exponent coincides with some notion of dimension. This has been demonstrated for several example systems as well as experimental data. In turn, we have found that in other cases the value of $\gamma$ drops to $1$. Further studies are necessary in order to better understand this complex relationship between power-law degree distributions and fractal scaling (i.e.\,under which general conditions related to a system's structural organisation both scaling exponent and fractal dimension coincide), particularly in continuous dynamical systems.

From a conceptual perspective, we would like to remark that studying a single
scalar property like the scaling exponent of a recurrence network or the fractal
dimension cannot provide a complete view on the structural organisation of a
nonlinear complex system. Specifically, both characteristics capture distinct
and complementary features related to the probability density of the invariant
measure. In this spirit, the power-law exponent $\gamma$ quantifies a fundamental 
property that has not been explicitly studied so far. Because its relationship
with the features of possible singularities of the invariant density is
intuitive (i.e.\,the emergence of power-law degree distributions has some clear physical meaning), one particular strength of studying the degree distribution of recurrence networks is that it potentially allows identifying the presence of such singularities in complex situations (\emph{e.g.,} for observational data).


\section{Acknowledgements} This work was partially supported by the German BMBF and the Leibniz association (projects PROGRESS and ECONS) as well as the German National
Academic Foundation. We acknowledge constructive comments by C.~Grebogi and M.~Zaks.

\bibliographystyle{eplbib}
\bibliography{zouetalbib_v4}

\end{document}